\documentclass[prl,twocolumn,aps,showpacs]{revtex4}
\include{psfig}

\begin{document}

\title{\bf Coherent current transport in wide ballistic Josephson junctions}
\author{P. Samuelsson$^a$, \AA. Ingerman$^b$}
\thanks{Present adress: Department of Physics, University of the Western Cape,
Cape Town, Bellville 7535, South Africa.}
\author{G. Johansson$^c$, E.V. Bezuglyi$^d$, V.S. Shumeiko$^b$, and
  G.  Wendin$^b$}
\affiliation{$^a$D\'epartement de Physique Theorique, Universit\'e de
Gen\`eve, CH-1211 Gen\'eve 4, Switzerland. \\
$^b$Department of Microtechnology and Nanoscience,
  Chalmers University of Technology, S-41296 G\"{o}teborg, Sweden. \\
$^c$Institut f\"ur Theoretische Festk\"orperphysik, Universit\"at Karlsruhe, 76128 Karlsruhe, Germany. \\
$^d$B. Verkin Institute for Low Temperature Physics and Engineering,
Kharkov 61164, Ukraine.}
\author{R. K\"ursten, A. Richter}
\thanks{Present adress: NTT Basic Research Labs., 3-1, Morinosato
Wakamiya, Atsugi, Kanagawa, 243-0198, Japan.}
\author{T. Matsuyama and U. Merkt}
\affiliation{Institut f\"ur Angewandte Physik, Universit\"at
Hamburg, Jungiusstrasse 11, D-20355, Germany.}
\pacs{74.45.+c,74.50.+r,73.23.Ad}

\begin{abstract}
We present an experimental and theoretical investigation of coherent
current transport in wide ballistic superconductor-two dimensional
electron gas-superconductor junctions. It is found experimentally that
upon increasing the junction length, the subharmonic gap structure in
the current-voltage characteristics is shifted to lower voltages, and
the excess current at voltages much larger than the superconducting
gap decreases. Applying a theory of coherent multiple Andreev
reflection, we show that these observations can be explained in terms
of transport through Andreev resonances.
\vspace{0.5cm}
\end{abstract}

\maketitle 

Ballistic superconducting junctions are of great interest for studying
fundamental properties of coherent electronic transport. The concept of Josephson
effect in superconductor-normal metal-superconductor junctions was
originally formulated for ballistic junctions \cite{Kulik}. However,
until recently, experiments could only be performed on metallic diffusive
junctions, which has limited the range of phenomena
observed. Experimental realization of the ballistic regime became
possible when a high-mobility two-dimensional electron gas (2DEG) was
employed to connect superconducting (S) electrodes\ \cite{Takayanagi1}. The recent
experimental interest \cite{Schapers} has been focused on junctions
with InAs 2DEGs which form highly transparent 2DEG-S interfaces with
large probability of Andreev reflection. Moreover, electrostatic
gating \cite{Takayanagi1,Takayanagi2} has made it possible to control junction
parameters.

The dc Josephson effect in ballistic S-2DEG-S junctions has been
extensively studied both experimentally
\cite{submicron,Takayanagi1,Schapers,Takayanagi2,dcjosephson} and theoretically
\cite{Kresin}. Less investigated are the transport properties of the
junctions in the presence of a voltage bias between the
superconductors, the regime of multiple Andreev reflection (MAR)
\cite{Klapwijk}. In particular, a conclusive experimental picture of
the subharmonic gap structure (SGS) in the current-voltage
characteristics is lacking. Moreover, the main tool for analyzing the
experimentally observed current-voltage characteristics has been the
theory of Octavio {\it et al}.\ \cite{OTBK} (OTBK). However, this is
not appropriate for junctions which show a dc Josephson effect,
i.e. which are in the coherent transport regime, since the OTBK theory
is restricted to incoherent MAR transport.

The theory for coherent MAR transport in junctions much shorter than
the superconducting coherence length \cite{marusual} has with great
accuracy explained experiments with atomic point contacts
\cite{marqpc}, both the details of the SGS at $eV = 2\Delta/n$ and the
magnitude of the excess current at voltages much larger than the
superconducting gap. This theory however cannot be directly applied to
the experimentally studied junctions
\cite{submicron,Takayanagi1,Schapers,Takayanagi2,dcjosephson}, generally of lengths
$L$ of the order of the superconducting coherence length $\xi_0$.

In this letter we report a systematic experimental investigation of
coherent current transport in ballistic Nb-InAs-Nb juctions with
2DEG-S interfaces of high transparency. For every junction we
investigated the three key transport characteristics: SGS, excess
current, and critical Josephson current. All investigated junctions
show a dc Josephson effect. Our main experimental observation is that
the SGS is shifted down in voltage upon increasing the distance
between the superconducting electrodes, i.e. the junction
length. Moreover, the excess current is found to decrease with
increasing junction length.

These observations can be explained within a coherent, multimode
MAR-theory. The shift of the SGS results from transport through
Andreev resonances \cite{Johansson,Ingerman}: the Andreev resonances,
closely related to the Andreev levels carrying the dc Josephson
current, are shifted to lower energies upon increasing the junction
length, consequently shifting the SGS down in voltage. Similarly, the
excess current decreases with increasing junction length for the
experimentally relevant junction lengths, $L \alt \xi_0$. 

\begin{figure}[h]
\centerline{\psfig{figure=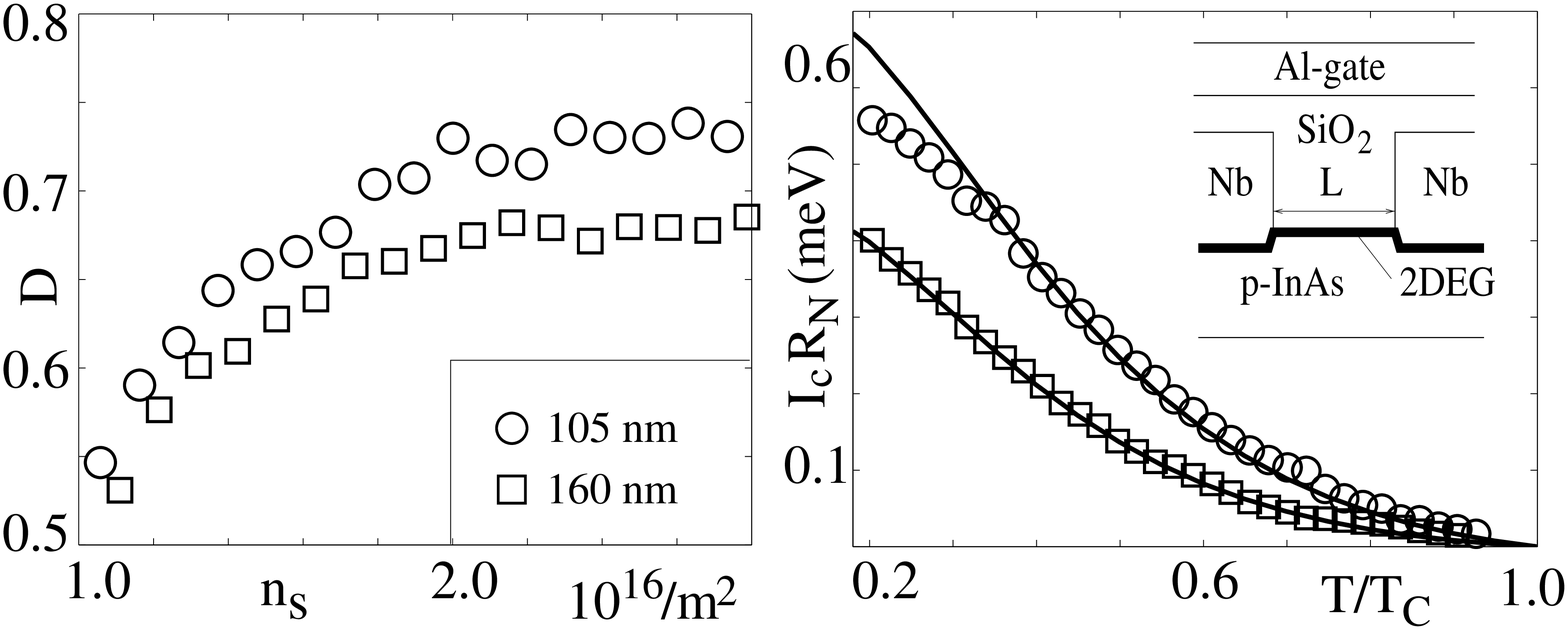,width=8.5cm}}
\caption{Left: Average normal transparency
$D=R_{\text{N}}/R_{\text{Sh}}$ as a function of $n_{\text{s}}$ for
samples of lengths $L=105$~nm and $160$~nm. Right:
$I_{\text{c}}R_{\text{N}}$ product as a function of $T$ for the same
samples. The thin lines are the theory, with
$I_{\text{c}}R_{\text{N}} (T=0)$ scaled for best fit by $1.65$
($L=105$~nm) and $2.4$ ($L=160$~nm) and $\hbar
v_{\text{F}}/E_{\text{A}}=0.54 \xi_{\text{0}}$ (see text). Inset: A
schematic cut through the junction, displaying the kink in the 2DEG.}
\label{fig1}
\end{figure}

A schematic picture of the junction is shown in Fig. \ref{fig1}. Two
niobium electrodes are deposited directly on the 2DEG which forms
spontaneously at the surface of single crystal InAs \cite{details}. In
the following step, the junction is covered by a SiO$_2$ layer that
provides insulation to the Al gate electrode grown on top. On three
different chips A,B,C, junctions of five different interelectrode
distances $L=105(\text{A}),115(\text{B}),145(\text{B}),160(\text{A})$
and $200(\text{C})$~nm have been fabricated. All junctions have a
width $W=50~\mu$m. The sheet carrier density $n_{\text{s}}$, the
effective mass $m^*$ and the mobility $\mu$ are controlled with the
top gate to give the largest 2DEG-conductance.  Typical parameters for
the unprocessed InAs are $n_{\text{s}}=2\times 10^{16}$ m$^{-2}$ and
$m^*=0.033 m_{\text{e}}$, where $m_{\text{e}}$ is the free electron
mass. This gives a Fermi wave length $\lambda_{\text{F}}=18$~nm and
$v_{\text{F}}=1.2\times10^6$ ms$^{-1}$. The mobility was not measured
on the actual samples, but is typically of order $\mu=10000$
cm$^2$(Vs)$^{-1}$. The measured critical temperature of Nb
$T_{\text{c}}=9.0$~K corresponds to a superconducting gap
$\Delta=1.4$~meV close to the bulk value. The superconducting
coherence length is thus $\xi_0=\hbar v_{\text{F}}/\Delta=600$~nm.
The normal resistance $R_{\text{N}}$ was measured at a voltage $eV \gg
\Delta$. The ratio between the Sharvin resistance
$R_{\text{Sh}}=h/(2e^2)[\lambda_{\text{F}}/2W]$ and the normal resistance, i.e. the
average transmission probability per conduction mode, is $D \sim 0.7$
for $n_{\text{s}}>2\times 10^{16}$ m$^{-2}$ (see Fig. \ref{fig1}). The
interfaces between the 2DEG and the metal electrodes are thus highly
transparent. Attributing the small difference in resistance between
the $105$ and $160$~nm junction to impurity scattering gives a mean
free path $l\approx 300$~nm. This demonstrates that the transport is
in the ballistic regime ($L<l$). All junctions show a dc Josephson
current, with $I_{\text{c}}R_{\text{N}}$ products $0.6,0.55,0.52,0.45$
at $1.8$~K monotonically decreasing with $L$ from $L=105$ to
$200$~nm. The temperature dependence of the dc Josephson current is
shown in Fig. \ref{fig1} for two junctions $L=105$ and $160$~nm.

\begin{figure}[h]
\centerline{\psfig{figure=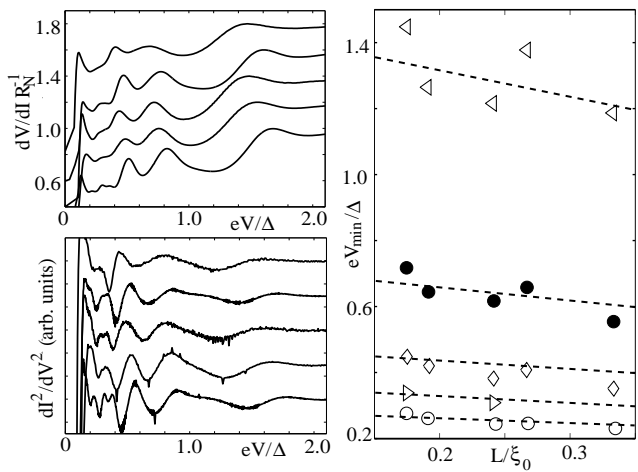,width=8.5cm}}
\caption{Left: Measured differential resistances $dV/dI$ (upper) and
(numerically differentiated) $d^2I/dV^2$ (lower) as a function of $V$
for $T=1.8$~K. The junction lengths are from bottom to top,
$L=105,115,145,160$ and $200$~nm. The traces have been successively
shifted for clarity. Right: Positions of the minima $V_{\text{min}}$
of the $d^2I/dV^2$ (lower left) for the SGS $n=1$ ($\triangleleft$),
$n=2$ ($\bullet$), $n=3$ ($\diamond$), $n=4$ ($\triangleright$, only
two minima clearly visible) and $n=5$ ($\circ$). Dashed lines are
given by Eq. (\ref{vmin}), with an additional length $\hbar
v_{\text{F}}/E_{\text{A}}=0.6\xi_{\text{0}}$.}
\label{fig2}
\end{figure}

In Fig. \ref{fig2} the experimentally measured differential resistance
for the different junctions is depicted. For all junctions, the
resistance shows SGS at voltages $eV<2\Delta$. As an overall tendency,
the SGS is shifted to lower voltages when increasing junction
length. For junctions on the same chip this holds strictly. All
SGS-features are symmetric around $V=0$ and scale roughly with
temperature like the gap $\Delta(T)$, up to $T_{\text{c}}$.

A decrease in the excess current as a function of length $L$ is also
observed in the experiment. The $I_{\text{exc}}R_{\text{N}}$ product
for the different junctions is shown in Fig. \ref{fig4} below,
displaying an overall decrease with increasing $L$. This holds
strictly, just as the shift of the SGS, for junctions on the same
chip. The systematic shift of the SGS and the decrease of the excess
current with junction length are the main experimental observations in
this work.

Based on these findings, we model the junction as a ballistic,
two-dimensional normal conductor of length $L$ and width $W$ connected
to two bulk superconducting electrodes via highly transparent
interfaces. The normal reflection in the junction is assumed to be
concentrated to the kink in the 2DEG formed during the junction
processing, see Fig. \ref{fig1}. For simplicity, the normal reflection
is taken to be specular, reducing the problem to summing over
independent modes. The width of the junction $W \gg
\lambda_{\text{F}}$, giving a number of transport modes
$2W/\lambda_{\text{F}}=5600$. No sources of decoherence,
e.g. inelastic scattering \cite{inelastic}, are taken into account in
the model.

We perform our theoretical analysis within the scattering approach to
the Bogoliubov-de Gennes equation \cite{Datta}. This allows us to
express the dc Josephson current \cite{Brouwer96} as well as the
current in voltage biased junctions \cite{Johansson} in terms of the
transmission and reflection amplitudes $t_m(E)$ and $r_m(E)$ of each
transport mode $m$ and the mode independent Andreev reflection
probability $\alpha(E)$ at the 2DEG-S interface. The scattering
amplitudes are given by $r_m=i\sqrt{R}(1+q_m^2)/(1+Rq_m^2)$ and
$t_m=q_m(1-R)/(1+Rq_m^2)$ where $q_m=\mbox{exp}[ik_mL]$;
$k_m(E)=[2m^*/\hbar^2(E_{\text{F}}-E_m+E)]^{1/2}$, where
$E_{\text{F}}$ is the Fermi energy and $E_m$ is the transverse mode
energy. The reflection probability of each interface $R$, taken to be
mode independent, is given from the averaged total transmission
probability $D$ as $R=(1-D)/(1+D)$, giving $R=0.16$ for $D=0.72$ ($D$
for the $105$~nm junction at $n_{\text{s}}=2\times 10^{16}$ m$^{-2}$).

The Andreev reflection amplitude at the 2DEG-S interface is given by
$\alpha(E)=\mbox{exp}[-i\mbox{arccos}(E/\Delta)]$. It is however not
possible to apriori conclude that quasiparticles are Andreev reflected
directly at the 2DEG-S interface near the kink, they might spend a
certain time $\hbar/E_{\text{A}}$ underneath the superconductor
\cite{dwelltime}. This can be incorporated into the model in the
simplest possible way by multiplying $\alpha(E)$ with a phase factor
$\mbox{exp}(iE/E_{\text{A}})$, where $E_{\text{A}}$ has to be
determined by comparison to the experimental data. \cite{Aminov}

The total dc-current can be written as a sum of the currents for the
individual modes \cite{summodes}, i.e. $I=\sum_mI_m$. In the general
case, it is not possible to analytically calculate the current for
arbitrary voltage, but one has to resort to numerics. It is important
to note that most transport modes $m$ have effective lengths
$L/\sqrt{1-(m/N)^2}$ close to the physical length $L$. Thus, the sharp
features in the SGS predicted in single mode junctions of finite
length \cite{Ingerman} are not washed out but are merely broadened by
the summation over the modes.

\begin{figure}[h]
\centerline{\psfig{figure=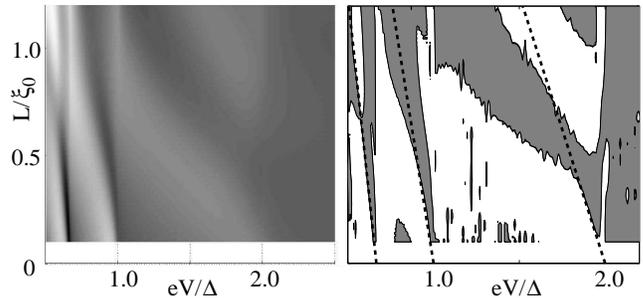,width=8.5cm}}
\caption{Left: Plot of the numerically calculated $dI/dV$ as function
of $eV/\Delta$ and $L/\xi_0$ for $R=0.16$ and $kT \ll \Delta$; light
areas correspond to peaks, dark to dips.  ($\hbar
v_{\text{F}}/E_{\text{A}}=0$). Right: A contour plot of the third
derivative $dI^3/dV^3$ around zero, giving the extremal points of
$dI^2/dV^2$ in the intersections of the dark and white areas. The
minima in Eq. (\ref{vmin}) are indicated by dashed lines.}
\label{fig3}
\end{figure}

The qualitative behavior of the SGS as a function of junction length
is clear from the numerics in Fig. \ref{fig3}. One sees how the SGS
are shifted down to lower voltages when increasing the junction length
$L$. The nature of this effect is the shift of broadened MAR
resonances\ \cite{Johansson,Ingerman}. The MAR resonances are closely
related to the Andreev levels carrying the dc Josephson current. The
Andreev levels are shifted towards lower energies upon increasing the
junction length, leading to a corresponding shift in the
SGS. Additional numerics shows that the general tendency of a shift of
the SGS towards lower voltages is independent of the reflection
probability $R$, while the details of the SGS depend on $R$. For
perfectly transparent 2DEG-S interfaces, however, the SGS are absent
for $kT \ll \Delta$\ \cite{Klapwijk} for any junction length
\cite{Ingerman}.

Quantitatively, we have performed a detailed numerical investigation
of the SGS $(n=1-5)$ in the experimentally relevant limit of small
normal reflectivity, $R \leq 0.16$ and small effective junction
length, $L_{\text{eff}}<\xi_{\text{0}}$. The effective length
$L_{\text{eff}} = L + \hbar v_{\text{F}}/E_{\text{A}}$ is the sum of
the geometric length of the junction $L$ and the distance $\hbar
v_{\text{F}}/E_{\text{A}}$ the quasiparticles propagate under the
superconductors before they are Andreev reflected. The analysis of the
numerical results shows that close to $V=2\Delta/en$ the second
derivative \cite{dercom} $d^2I/dV^2$ has minima which depend on
junction length as
\begin{equation}
V_{\text{min}}=\frac{2\Delta}{en}\left(1-0.4\frac{L_{\text{eff}}}{\xi_{\text{0}}}
\right),
\hspace{0.5cm} n=1,2,..
\label{vmin}
\end{equation}
for $L_{\text{eff}}<0.4\xi_{\text{0}}$ for all $n$ (for $n>1$, it
holds for even longer $L_{\text{eff}}$). This length dependence is in
strong contrast to the result of the theory of incoherent MAR
\cite{OTBK}, which predicts dips in the second derivative at voltages
$V_{\text{min}}=2\Delta/en$ {\it independent} of junction length
\cite{Zimmermann}.

Comparing theory and experiment, we see that the coherent MAR theory
(Fig. \ref{fig3}) is able to explain the overall shift in the
experimentally observed SGS towards lower voltage for increasing
junction length $L$ (see Fig. \ref{fig2}). However, to fit the
measured positions of the dips in the second derivative with
Eq. (\ref{vmin}), one needs to include an additional length $\hbar
v_{\text{F}}/E_{\text{A}}$. The best fit, including the SGS to fifth
order, is obtained for $\hbar
v_{\text{F}}/E_{\text{A}}=0.54\xi_{\text{0}}$ for both junctions on
chip A, $0.7\xi_{\text{0}}$ for both junctions on chip B and
$0.7\xi_{\text{0}}$ for chip C. We note that the values for the
effective lengths $L_{\text{eff}}$ are partly outside the range of
validity of Eq. (1), which might introduce some error in the fitted
values $\hbar v_{\text{F}}/E_{\text{A}}$. The fact that the junctions on the
same chip give the same $E_{\text{A}}$ suggests that $E_{\text{A}}$ is
sensitive to the sample preparation.

Although the dip and peak positions of the SGS can be well reproduced
by the theory, the theoretically calculated amplitudes (not shown) of
the SGS-features are much larger than the ones observed in the
experiment. One reason might be residual inelastic scattering or
dephasing, not taken into account in the model. Another reason might
be that the theory assumes a voltage bias, while the measurements are
carried out with current bias, giving rise to voltage fluctuations
which smear the SGS.
\begin{figure}[h]
\centerline{\psfig{figure=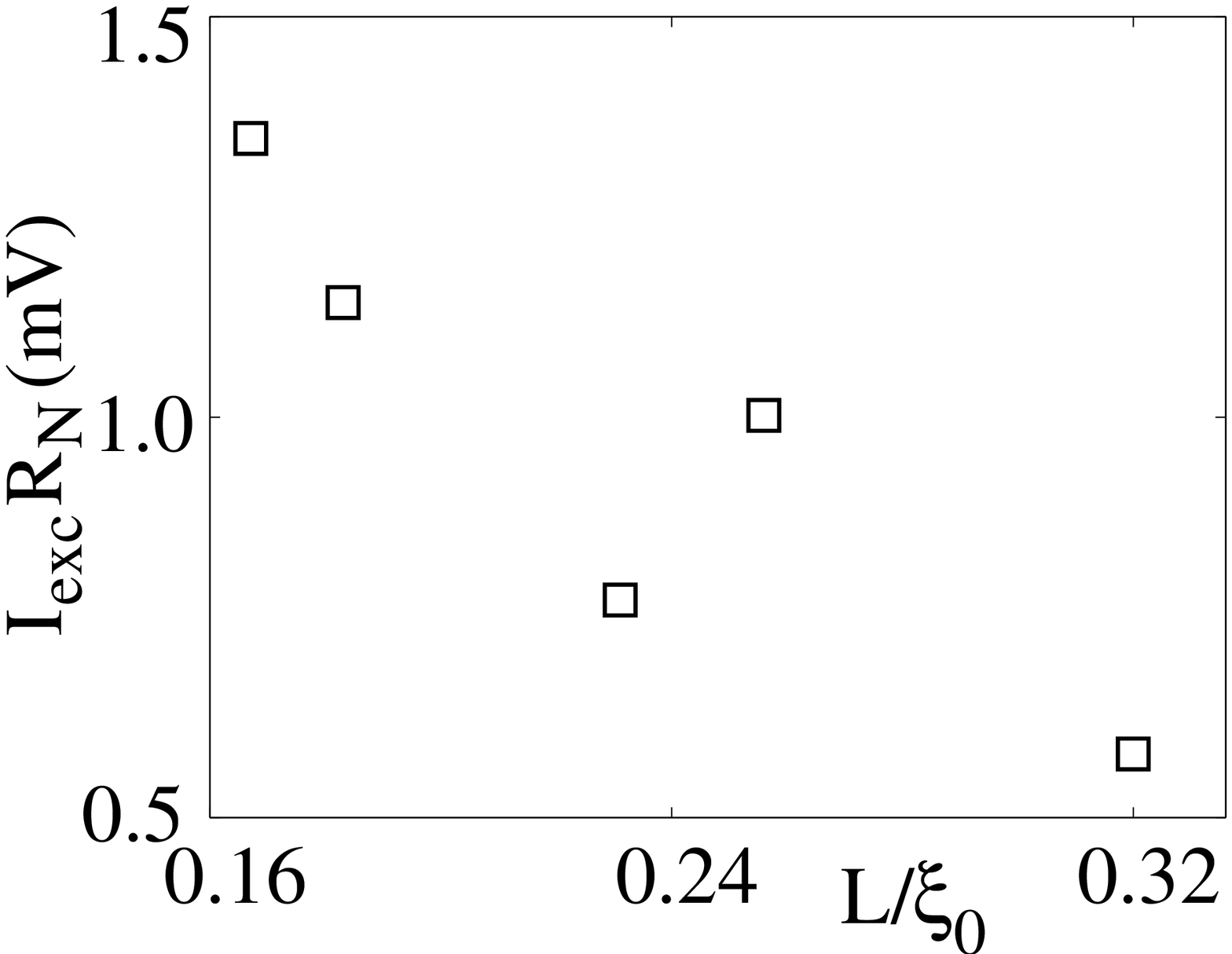,width=4.25cm}\psfig{figure=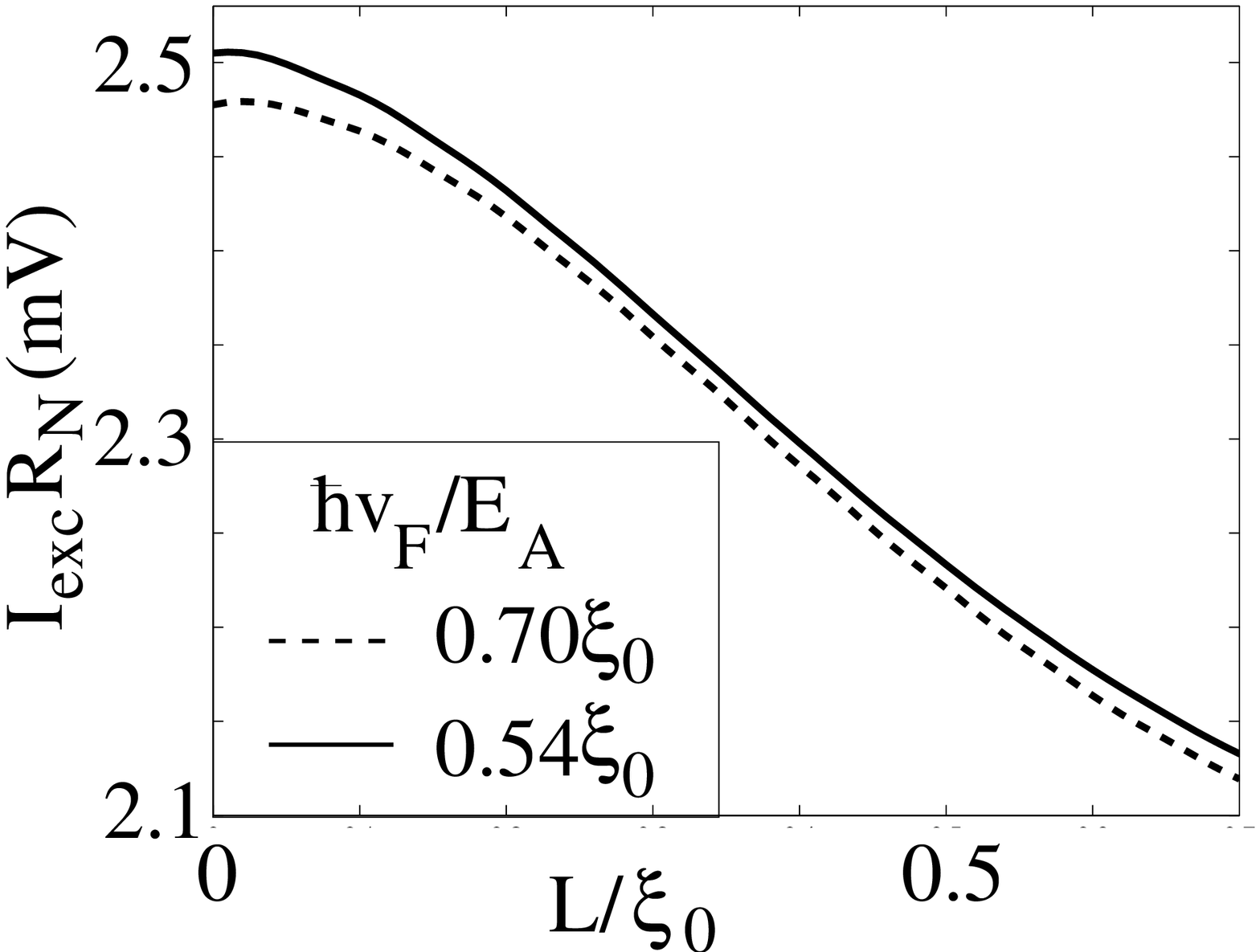,width=4.25cm}}
\caption{Left: Excess current at $T=1.8$~K for the five experimentally
investigated junctions. Right: Numerically calculated excess current
as a function of length $L$, for $R=0.16$ and $k_BT\ll\Delta$.}
\label{fig4}
\end{figure} 

The experimentally measured excess current
$I_{\text{exc}}=I-V/R_{\text{N}}$ at voltages $V\gg \Delta/e$ is
presented in the left panel in Fig. \ref{fig4}. As seen from numerics
in the right panel of Fig. \ref{fig4}, for both
$E_{\text{A}}=1.4\Delta$ and $1.8\Delta$, the excess current decreases
monotonically with junction length $L$ for all lengths
$L<\xi_{\text{0}}$, in qualitative agreement with the
experiment. However, this decay is slower than the measured
one. Moreover, the theoretical excess current exceeds the experimental
one. This discrepancy can be attributed to residual inelastic
scattering, similar to the suppression of the amplitude of the SGS
mentioned above. The excess current is carried by quasiparticles which
traverse the junction one and two times. The single-particle processes
give a negative contribution to the excess current while the
two-particle processes yield a positive contribution. Since
quasiparticles carrying the two-particle current spend longer time in
the junction, inelastic scattering leads to a stronger suppression of
the two-particle current compared to the one-particle current. This
results in an excess current with a stronger length-dependent decay
than predicted by the coherent theory.

We also compare the experimentally measured dc Josephson currents with
the result of the model. The expression for Josephson current in terms
of the normal scattering amplitudes and the Andreev reflection
amplitudes is given in Ref. \cite{Brouwer96}. Using the value of
$E_{\text{A}}$ deduced from the SGS fitting, we see in Fig. \ref{fig1}
that the temperature dependence of the experimentally measured
critical current is well reproduced by the theory.  However, the
theoretical overall magnitude is $1.65$ times (for $L=105$~nm) resp.
$2.4$ times (for $L=160$~nm) larger than the experimental value. This
discrepancy indicates, in line with the discussion for the
SGS-amplitude and the excess current above, the presence of some
decoherenece mechanism. Note that the discrepancy is much smaller than
previously reported \cite{submicron,Takayanagi1,Takayanagi2}, because the longer
effective junction length reduces the theoretical critical current to
a value more similar to the experimental one. The significant part is
that this effective length has been determined from a best fitting to
the SGS.

In conclusion, we have studied, experimentally and theoretically,
coherent current transport in wide ballistic S-2DEG-S junctions. It is
found experimentally that both the SGS and the excess current show a
systematic dependence on length of the junction. We show that these
observations can be qualitatively explained within a coherent theory
of MAR. Our investigation also points towards additional mechanisms,
e.g. decoherence, to be included in the theory to obtain a
quantitative agreement with experiments.

The work was supported by MANEP (P.S.), STINT (\AA.I.), the Humboldt
Foundation, the BMBF and the ZIP-programme of the German government
(G.J.), the Swedish grant agencies TFR (P.S.), SSF (V.S.), VR (G.W.,
V.S.) and KVA (E.B.), and by the DFG via SFB 508 (R.K., A.R., T.M. and
U.M.)


\begin{thebibliography}{999}
\bibitem{details}
Details of the junction fabrication can be found in A. Chrestin et
al., \cite{Schapers}.
\bibitem{dercom} 
We focus on the minima in $d^2I/dV^2$ which, in
contrast to the extremal points in $dI/dV$, approach $2\Delta/en$ for
$L\rightarrow 0$ and are essentially independent on $R$ for $R<0.16$.
\end{thebibliography}
\end{document}